\documentclass[preprint,showpacs,preprintnumbers,amsmath,amssymb,prl,nofootinbib]{revtex4-1}
\usepackage[dvips]{graphicx}
\newcommand{\be}{\begin{equation}}\newcommand{\ee}{\end{equation}}
\newcommand{\bea}{\begin{eqnarray}}\newcommand{\eea}{\end{eqnarray}}
\newcommand{\bi}{\begin{enumerate}}
\newcommand{\ei}{\end{enumerate}}
\newcommand{\bref}[1]{(\ref{#1})}
\newcommand{\nn}{\nonumber}
\newcommand{\A}{\alpha}\newcommand{\B}{\beta} \newcommand{\gam}{\gamma}
\newcommand{\G}{\gamma}\newcommand{\D}{\delta} 
\newcommand{\ep}{\epsilon}

\newcommand{\s}{\sigma}
          
\newcommand{\z}{\zeta}          
\newcommand{\h}{\eta}

            \newcommand{\Sig}{\Sigma}

\newcommand{\bS}{\Sigma}%{{\bf{\Sigma}}}
\newcommand{\ba}{\overline }
\def\6{\partial}\def\7{\tilde}\def\8{\hat}
\def\pa{\partial}

\def\CC{{\cal C}}\def\CG{{\cal G}}\def\CL{{\cal L}}
\def\CA{{\cal A}}\def\CF{{\cal F}}
\def\CM{{\cal M}}

\def\={{\;=\;}}
\def\vs{\vskip 4mm}\def\+{{\;+\;}}
\def\SM{{superMaxwell }}
\newcommand{\T}{{\theta}}%{{\theta \hskip-2.03mm \theta}}

\newcommand{\bQ}{Q}%{{\bf{Q}}}
\newcommand{\bph}{{\phi}}% {{\phi \hskip-2.2mm \phi}}

\newcommand{\blam}{{\lambda }}%{{\lam \hskip-2.2mm \lam}}

\newcommand{\bA}{A}%{{\mathbb{A}}} %%%{{ A \hskip-2.85mm {A}}}
\newcommand{\bD}{D} %{{\mathbb{D}}} %%{{\bf{D}}}
\newcommand{\bF}{F} %{{\mathbb{F}}} %%%{{ F \hskip-3.05mm {F}}}
\newcommand{\rD}{{\rm D}}%%{{\mathbb{D}}}
\newcommand{\bW}{W}%{{\mathbb{W}}} %%%{{ W \hskip-4.7mm {W}}}

\def\c99{ {i}}%

\newcommand\half{{\frac{1}{2}}}
\def\phiD{{\phi}}%{{\varphi}}
\begin{document}
\preprint{ICCUB-09-300,UB-ECM-PF 09/29,Toho-CP-0991}
% title changed by editorial reason \title{SuperMaxwell Algebra and Superparticle in
%Constant SUSY Gauge Backgrounds}
\title{Maxwell Superalgebra and Superparticle in
Constant Gauge Backgrounds}
\author{Sotirios Bonanos$^1$}\author{Joaquim Gomis$^{2,3}$}
\author{Kiyoshi Kamimura${}^4$}\author{Jerzy Lukierski${}^5$}
\affiliation{${}^1$Institute of Nuclear Physics, NCSR Demokritos,
15310 Aghia Paraskevi, Attiki, Greece}
\affiliation{${}^2$Departament ECM and ICCUB, Universitat de Barcelona, 
Diagonal 647, 08028 Barcelona, Spain}
\affiliation{${}^3$Facultad de F\'{\i}sica, Universidad Cat\'olica de Chile,
Santiago 22, Chile}
\affiliation{${}^4$Department of Physics, Toho University Funabashi  
274-8510,  Japan }%
\affiliation{${}^5$Institute of Theoretical
Physics, Wroclaw University, pl. Maxa Borna 9, 50-204 Wroclaw, Poland}

\date{\today}
\begin{abstract}
We present  the \SM  algebra: an $N$=1, $D$=4 algebra with two Majorana supercharges, obtained as the minimal enlargement of 
superPoincar\'{e} containing the Maxwell algebra as a subalgebra.
The new superalgebra describes the supersymmetries of generalized  $N$=1, $D$=4 superspace in the presence of  a  constant Abelian SUSY field strength background. Applying the techniques of non-linear coset realization to the Maxwell supergroup we propose a new $\kappa$-invariant massless superparticle model providing a dynamical realization of the \SM  algebra.
\end{abstract}
\maketitle

%%%%%%%%%%%%%%%%%%%%%%%%%%%%%%%%%%%%%%%%%%%%%%%%%%%%%%%%%%%%%%%%%%
\section*{Introduction}

Recently, after the discovery of cosmic microwave background (CMB) and the 
mystery of dark energy \cite{Frieman:2008sn}, it appears interesting to 
consider some field densities uniformly filling 
space-time. One such modification of empty Minkowski space is obtained by 
adding a constant electromagnetic (EM) field background,  parametrized by 
the additional field  degree of freedom
$f_{\mu\nu}$. The presence of a constant EM field 
modifies the Poincar\'{e} symmetries into the so-called Maxwell symmetries
 \cite{Bacry:1970ye}-\cite{Gibbons:2009me}. 
The difference from the  
Poincar\'{e} algebra consists in the de-Sitter-like substitution
\footnotemark[{1}]
\footnotetext[{1}]{The Bacry-Combe-Richard(BCR) algebra \cite{Bacry:1970ye} is 
a subalgebra of the Maxwell algebra in which  $Z_{\mu\nu}$ takes fixed numerical values.
}\footnotemark[2]
\footnotetext[2]{
Recall that dark energy is sometimes described by the addition of a cosmological term, or replacement of ``empty'' Minkowski space by de-Sitter space.}
\begin{equation}
\label{eqjap1.1}
[P_\mu, P_\nu ] =i Z_{\mu\nu} \,.
\end{equation}
The additional tensorial generators $Z_{\mu \nu}$ are however Abelian
and satisfy the relations
\begin{eqnarray}
\label{eqjap1.2}
[M_{\mu \nu}, Z_{\rho \tau}] & = &
- i (\eta_{\nu\rho} Z_{\mu\tau} - \eta_{\nu \tau} Z_{\mu \rho}
+ \eta_{\mu \tau} Z_{\nu \rho} - \eta_{\mu \rho} Z_{\nu \tau}),
\cr
[P_\mu , Z_{\nu \rho} ] & = & 0, \qquad \qquad
[Z_{\mu \nu} , Z_{ \rho \tau} ] = 0 \, .
\end{eqnarray}

In the same way as the Poincar\'{e} algebra is the $R \to \infty$ limit
($R = dS$ radius) of de-Sitter 
algebra, the Maxwell algebra  $\CM_4$=$(M_{\mu\nu}, P_\mu , Z_{\mu\nu}$)
given in \bref{eqjap1.1}-\bref{eqjap1.2} can be obtained by a suitable contraction of the de-Sitter algebra $(\widetilde{M}_{\mu\nu}, P_\mu)$ enlarged in a semisimple way by the Lorentz generators ${M}_{\mu\nu}$
(see also \cite{Gomis:2009vm}). 
Performing the rescaling $ 
P_\mu\to{\A^{-1}}P_\mu,\;
\widetilde{M}_{\mu\nu} \to {\alpha^{-2}}Z_{\mu\nu},\;
{M}_{\mu\nu}\to {M}_{\mu\nu}
$ 
one obtains in the limit $\alpha \to 0$ the Maxwell algebra $\CM_4$.

In order to interpret the Maxwell algebra and the corresponding Maxwell group, 
a Maxwell group-invariant particle model on the extended space-time 
$(x^{\mu}, \phi^{\mu\nu})$ with the translations of $\phi^{\mu\nu}$ 
generated by
$Z_{\mu\nu}$ has been studied \cite{Bonanos:2008kr}-\cite{Gibbons:2009me}. 
 The interaction term described by a Maxwell-invariant one-form introduces new 
tensor degrees of freedom $f_{\mu\nu}$ - momenta conjugate to $\phi^{\mu\nu}$. 
In the equations of motion they play the role of a  background EM field which is 
constant on-shell and leads to a closed, Maxwell-invariant two-form.

The aim of this paper is to obtain the supersymmetric extension of the Maxwell 
symmetries with  new N=1 \SM algebra and to investigate the corresponding 
superMaxwell-invariant massless superparticle model. 
 (For massive superparticles one has to consider the N=2 
supersymmetries in D=4 \cite{de Azcarraga:1982dw}.) Analogously to the 
Maxwell case, one can introduce the generalized  phase space  with coordinates
 $(x^\mu, \theta^\alpha, \phi^{\mu\nu}, \phi^\alpha, \phiD)$ and conjugate 
momenta  $(p_\mu, \zeta_\alpha, f_{\mu\nu}, \tilde\lambda_\alpha, \rD)$.
Since $( \phi^{\mu\nu}, \phi^\alpha, \phiD)$ are cyclic coordinates the 
conjugate momenta $(f_{\mu\nu}, \tilde\lambda_\alpha, \rD)$
are constant on-shell describing the
 constant Abelian SUSY  N=1 gauge field background.
In this way one gets the massless superparticle 
interacting with $x$-independent  field strength superfield $W_\alpha (\theta)$
 \begin{equation}
 \label{eqjap1.4}
 \bW_\alpha (\theta) = i\tilde\lambda_\alpha -
 \frac{i}{2} \, f_{\mu\nu} (\bar{\theta} \gamma^{\mu\nu})_\alpha
 - i \rD (\bar{\theta} \gamma_5)_\alpha.
 \end{equation}
 We see therefore that the \SM symmetries  describe the geometry of N=1 
superspace $(x^\mu, \theta^\A)$ in the presence of constant SUSY gauge field 
background $(f_{\mu\nu}, \tilde\lambda_\alpha, \rD)$.  It is also noted
that the  
superparticle model is invariant  under
$\kappa$-transformations, which eliminate half of the Grassmann
superspace coordinates $\theta^\alpha$.
%%%%%%%%%%%%%%%%%%%%%%%%%%%%%%%%%%%%%%%%%%%%%%%
\section*{Particle model with Maxwell  symmetry}
%%%%%%%%%%%%%%%%%%%%%%%%%%%%%%%%%%%%%%%%%%%%%%
To formulate {a} 
relativistic particle model, invariant under the Maxwell group,
it is convenient to use the nonlinear coset realizations method \cite{Coleman}.
The coset $G/H$=Maxwell/Lorentz which we employ is parametrized as in
 \cite{Bonanos:2008kr}-\cite{Gibbons:2009me}, 
$\;
g=e^{iP_\mu x^\mu}e^{\frac{i}{2}Z_{\mu\nu}\phi^{\mu\nu}}\; .
$ %\label{cosetg0}\ee
The basic Maurer-Cartan (MC) form is
\be
\Omega=-i g^{-1}dg= P_\mu L^\mu+\frac12
Z_{\mu\nu}L_Z^{\mu\nu}+\frac12 M_{\mu\nu}L_M^{\mu\nu}, \label{MCform}
\ee
{where}\be
L^\mu=dx^\mu, \;
L_Z^{{\mu\nu}}=d\phi^{\mu\nu}+\frac12(x^\mu dx^\nu-x^\nu dx^\mu), \;
L_M^{{\mu\nu}}=0. \label{explicit1}
\ee
The particle action invariant under the Maxwell algebra \bref{eqjap1.1}
and \bref{eqjap1.2} is described by the following Lagrangian:
\be
\CL={\frac{\dot x_\mu \dot{x}^\mu}{2e}
- \frac{m^2}{2}e}+\frac{1}2 f_{\mu\nu}L_Z^{\mu\nu*},
\label{lagrangian2}
\ee
where $e$ is the einbein 
implementing the diffeomorphism invariance, 
$f_{\mu\nu}$ is a tensorial variable canonically conjugate 
to the new coordinates $\phi^{\mu\nu}$
and $ L_Z^{\mu\nu*}$ is the pullback of $ L_Z^{\mu\nu}$. 
In the proper time gauge, one obtains from \bref{lagrangian2} 
the equations of motion
\be\label{eqjap9}
m \ddot{x}_{\mu}  =  f_{\mu\nu} \dot{x}^\nu,\quad 
\dot{f}_{\mu\nu} =  0, \quad
\dot{\phi}^{\mu\nu} = - \frac{1}{2}(x^\mu \dot{x}^\nu -x^\nu \dot{x}^\mu).
\ee
They describe the motion of a particle in a EM field $f_{\mu\nu}$,
 which is constant on-shell. 
The EM potential is described by the one-form $\CA=\half f_{\mu\nu} 
L^{\mu\nu}_{\rm Z}$. In the closed two-form field strength
 \begin{equation}
 \label{eqjap12}
\CF = d\CA = \half f_{\mu\nu} L^\mu \wedge L^\nu
 + \half d f_{\mu\nu} \wedge L^{\mu\nu}_{\rm Z}
 \end{equation}
the second term vanishes on-shell due to \bref{eqjap9} and the field strength components are constants $f_{\mu\nu}$.
%%%%%%%%%%%%%%%%%%%%%%%%%%%%%%%%%%%%%%%%%%%%%%%%%%%%%%%%%%%%%%%%%%%%%%%
\section*{From Maxwell algebra to \SM algebra}

We start with the following extension of the superPoincar\'{e} algebra in D=4
with Majorana supercharges $Q_\alpha$
as  ($\A,\B=1,2,3,4 $)
\be
\{{\bQ}_\A,{\bQ}_\B\}=2(C\G^{\mu})_{\A\B}P_\mu,
\quad \left[P_{\mu},~P_{\nu}\right]= i Z_{\mu\nu}. 
\label{13}
\ee
In order to verify the $(P,Q,Q)$ Jacobi identity, $P_\mu$ cannot commute with
$Q_\A$
but requires a new Majorana charge $\Sig_\A$
defined as
\be
\left[P_\mu,{\bQ}_\A\right]=-i\bS_{\B}(\G_{\mu}{)^{\B}}_\A.
\label{14}
\ee
One can show from Jacobi identities that
\be
\{{\bQ}_\A,{\bS}_{\B}\}=\frac{1}2(C\G^{\mu\nu})_{\A\B}\,Z_{\mu\nu}.
\label{15}
\ee
 $\Sig_\A$, as well as $Q_\A$, transforms as a spinor under Lorentz transformations,
\be
\left[M_{\rho\s},\bQ_{\A}\right]=-\frac{i}2(\bQ \G_{\rho\s})_{\A},\quad
\left[M_{\rho\s},\bS_{\A}\right]=-\frac{i}2(\bS \G_{\rho\s})_{\A}.
\label{2.2}
\ee
Together with relations \bref{eqjap1.1} and  \bref{eqjap1.2}
 the superalgebra $\CG$
=$(M_{\mu\nu}, P_\mu,Z_{\mu\nu},$ $
 Q_\alpha, \Sig_\alpha)$ is shown to close due to the gamma matrix identity
$ (C\G^\mu)_{(\A\B}(C\G_{\mu})_{\gam\D)}=0,\; (\A\B\gam\D\;
$%\label{pidentA}\ee
{symmetric sum)} valid in D=4.
$\CG$ defines the minimal \SM
algebra containing the Maxwell algebra $\CM_4$ as a subalgebra.

Consistently with the Jacobi relations one can also add
a scalar central charge $B$
in \bref{15} as
\be
\{{\bQ}_\A,{\bS}_{\B}\}=\frac12(C\G^{\mu\nu})_{\A\B}\,Z_{\mu\nu}\,+\, 
(C\G_5)_{\A\B}\,{B}
\label{2.6}\ee
 and obtain the centrally extended algebra
${\widetilde\CG}$=$( M_{\mu\nu}, P_\mu,  $ $ Z_{\mu\nu},Q_\alpha, 
\Sig_\alpha,B)$.
It can be shown that the central charge $B$ corresponds to the constant mode of
an auxiliary scalar in the  ``off-shell" supersymmetric U(1) gauge field 
theory.

Two Casimir operators  of the Maxwell algebra  obtained in \cite{Bacry:1970ye}-\cite{Schrader:1972zd}
\be
\CC_2=Z_{\mu\nu}Z^{\mu\nu},\quad \CC_3=Z_{\mu\nu}\7Z^{\mu\nu}, \quad
(\7Z^{\mu\nu}\equiv\frac12\ep^{\mu\nu\rho\s}Z_{\rho\s}) \ee
are also Casimir operators of the \SM algebra $\CG$ but the third 
mass Casimir operator requires a fermionic term
\be
\CC=P^2+M_{\mu\nu}Z^{\mu\nu}+{{i}}\bS C^{-1}\bQ.
\label{C1}\ee
For the centrally extended algebra $\widetilde\CG$ 
the Casimir $\CC$ ceases to commute with $Q$ and $\Sig$.
However, in the presence of an additional chiral symmetry charge
$B_5$ satisfying
\be
\left[B_5,{\bQ}_\A\right]=-i(\bQ\G_{5})_\A,
\quad
\left[B_5,{\bS}_\A\right]=i(\bS\G_{5})_\A,
\label{chiral}
\ee
we can construct the extension of Casimir $\CC$
\be
\widetilde \CC=P^2+M_{\mu\nu}Z^{\mu\nu}+{{i}}\bS C^{-1}\bQ-\,B_5\,B,
\label{CasC1}\ee
which becomes a Casimir of  the algebra $\CG_5$=$(M_{\mu\nu},P_\mu, $ $Z_{\mu\nu}, Q_\alpha, \Sig_\alpha,B,B_5)$. The super 
algebra $\CG_5$ will be
realized in a massless particle model in the next section.

%%%%%%%%%%%%%%%%%%%%%%%%%%%%%%%%%%%%%%%%%%%%%%%%%%%%%%
\section*{Massless Superparticle model with Maxwell supersymmetry}

We construct a massless superparticle model using a non-linear realization of
the \SM algebra $\CG_5$. The supergroup element $\widetilde{g}$ is 
parametrized as
\be
\widetilde{g}= e^{\frac{i}2Z_{\mu\nu}\phi^{\mu\nu}}\,
e^{iP_\mu x^\mu}\,  e^{i\bS_{\A}{\phi}^{\A}}\, e^{{i\bQ}_\A\T^\A}\,e^{i{B}\,\phiD}
\label{coset}\ee
using the supercoset $G/H$=$\CG_5/(M\times B_5)$ 
\cite{bonanos}.
Here the chiral generator $B_5$ is in the unbroken subgroup because we
construct a massless particle.
The components of the MC form
$\widetilde{\Omega} = -i \widetilde{g}^{-1} d \widetilde{g}$ are
\bea
\widetilde{L}^\mu&=&dx^\mu+{i}(\ba\T\G^\mu d\T),\quad \widetilde{L}^\A=d\T^\A,\quad
\widetilde{L}_M^{\mu\nu}=0,
\nn\\
\widetilde{L}_Z^{\mu\nu}&=&d\phi^{\mu\nu}+{i}(\ba\T\G^{\mu\nu})_\A d\bph^\A
+\frac{1}2(x^{\mu}dx^{\nu}-x^{\nu}dx^{\mu})\cr
&&+\frac{{i}}2\,(\ba\T\G^{\mu\nu}\G_{\rho}\T)
(dx^\rho+\frac{{i}}6(\ba\T\G^\rho d\T)),
\nn\\ 
\widetilde{L}_\Sig^{\A}&=&d\bph^{\A}+(\G_{\rho}\T)^\A
(dx^\rho+\frac{{i}}3(\ba\T\G^\rho d\T)),\quad
\widetilde{L}^5=0,
\nn\\
\widetilde{L}_B&=&d\phiD+{{i}}(\ba\T\G_{5})_\A d\bph^\A
+\frac{{i}}2\,(\ba\T\G_5\G_{\rho}\T)
(dx^\rho+\frac{{i}}6(\ba\T\G^\rho d\T))
\nn\\ \label{MCsol}\eea
and verify the corresponding MC equations
\bea
 d\widetilde{L}^\mu&=& {i}
{\overline {\widetilde{L}}}\G^{\mu}
\widetilde{L}-\widetilde{L}_M^{\mu\nu}
\widetilde{L}_\nu,
\quad
 d\widetilde{L}_M^{\mu\nu}=-
 \widetilde{L}_M^{\mu\rho}\h_{\rho\s}
 \widetilde{L}_M^{\s\nu},
\nn\\
 d\widetilde{L}_Z^{\mu\nu}&=&
 \widetilde{L}^\mu\,\widetilde{L}^\nu+ {i}{\overline { \widetilde{L}}}\G^{\mu\nu}
 \widetilde{L}_\Sig-
\widetilde{L}_M^{\mu\rho}\h_{\rho\s}\widetilde{L}_Z^{\s\nu}-
\widetilde{L}_Z^{\mu\rho}\h_{\rho\s}\widetilde{L}_M^{\s\nu}
,
\nn\\
 d
\widetilde{L}^\A&=&\,(\G_5 \widetilde{L})^\A\, \widetilde{L}^5\,-\,\frac14
\widetilde{L}_M^{\mu\nu}(\G_{\mu\nu}
\widetilde{L})^\A ,
\nn\\
 d\widetilde{L}_\Sig^{\A}&=&(\G_{\mu}
 \widetilde{L})^{\A}\, \widetilde{L}^\mu-
(\G_5 \widetilde{L}_\Sig)^\A\,
\widetilde{L}^5-\frac14
\widetilde{L}_M^{\mu\nu}(\G_{\mu\nu}\widetilde{L}_\Sig)^\A,
\nn\\
 d\widetilde{L}_B&=& {i}{\overline {\widetilde{L}}}\,\G_{5}\,
\widetilde{L}_\Sig,
\quad
 d\widetilde{L}^5=0.
\label{MCBS}\eea
These MC equations provide a dual formulation  of the \SM algebra 
introduced in the previous section.

The massless super particle action invariant under the \SM group is
\be
\CL=\frac{\pi_\mu^{2}}{2e}+\CL^{I*};\quad
\CL^I =\frac12f_{\mu\nu}\widetilde{L}_Z^{\mu\nu}
+{{i}}\blam_\A\widetilde{L}_\Sig^\A+ \rD \widetilde{L}_B
,\label{Lag0}
\ee
where $\pi_\mu=\dot x_\mu+i\ba\T\G_\mu\dot\T$ is the pullback of $\widetilde{L}_\mu$
to the world line and  $e$ describes the einbein.
Here   $(f_{\mu\nu},\blam_\A,\rD)$  are dynamical variables
transforming as  Lorentz tensor, Majorana spinor and scalar respectively.
The interaction Lagrangian can be written  explicitly as
\be
\CL^{I*}=
\frac12\,f_{\mu\nu}\,\dot\phi^{\mu\nu}+
{{i}}\7\blam_\A\,\dot\bph^{\A}+\rD\,\dot\phiD
+\pi^\mu\,A_\mu+\dot\T^\A\7\bA_\A,
\label{Lag021}
\ee where 
\be
\7\blam_{\A}=\blam_{\A}+\rD(\ba\T\G_{5})_\A+\frac12f_{\mu\nu}
(\ba\T\G^{\mu\nu})_\A
\label{deftlam}\ee
and the U(1) SUSY gauge potentials are
\bea
\7\bA_\A&=&{{i}}(\ba\T\G^\mu)_\A[-\frac12f_{\mu\nu}x^{\nu}+
{{i}}(\frac23\7\blam-\frac18\ba\T\G_{\rho\s}f^{\rho\s}{\hskip-1.5mm}
-\frac14\rD\ba\T\G_5)\G_\mu\T],
\nn \\ 
A_\mu&=&-\frac12f_{\mu\nu}x^{\nu}
+{{i}}(\7\blam-\frac14\ba\T\G_{\rho\s}f^{\rho\s}{\hskip-1.5mm}
-\frac12\rD\ba\T\G_5)\G_\mu\T. 
\label{U1A}\eea
The variation of $\CL$ with respect to $(\phi^{\mu\nu}, \phi^\alpha , \phiD)$ 
gives
\be
\dot f_{\mu\nu}\,=\dot{\7\blam}_\A\,=\dot\rD=0,
\label{feom}\ee
i.e.,  the U(1) superpotentials \bref{U1A} are functions of the superspace 
coordinates
$(x^\mu,\T^\A)$ and the variables  $(f_{\mu\nu},\tilde\lambda_\alpha, \rD)$ 
which take constant values on-shell.
The variation of $\CL$ with respect to $(f_{\mu\nu},\7\blam_\A,\rD)$ gives
 the equations for the variables $(\phi^{\mu\nu}, \phi^\alpha , \phiD)$
\bea
(\widetilde{L}_Z^{\mu\nu})^*=(\widetilde{L}_\Sig^{\A})^*=(\widetilde{L}_B)^*=0.
\label{phieom}\eea
The variation of $\CL$ with respect to  $e$
puts the momenta $\pi_\mu$ on mass shell with vanishing mass
\be
{\pi^2}=0.
\label{masslesspi}\ee
Finally, the variation of $\CL$ with respect to  $(x^\mu,\T^\A)$ gives, using
\bref{U1A}--\bref{feom}, the superparticle
 equations of motion in superspace,
\bea
 \frac{d}{d\tau} (\frac{\pi_\mu}{e})&=&\pi^\nu F_{\mu\nu}+\dot\T^\B 
\bF_{\mu\B},
\label{xeoms}\\
2{{i}}\,(\dot{\ba\T}\G^\mu)_\A\, (\frac{\pi_\mu}{e})
&=&\pi^\nu \bF_{\nu\A}\, ,
\label{Theom}\eea
where the superfield strengths are
($D_\alpha = \partial_\alpha + i (\bar{\theta}\gamma^\mu)_\alpha \partial_\mu$)
\bea    F_{\mu\nu}&=&(\pa_\mu A_\nu-\pa_\nu A_\mu) =f_{\mu\nu},\qquad \qquad
\nn\\
\bF_{\mu\A}&=&(\pa_\mu \7\bA_\A-\bD_\A A_\mu)=i(\blam\G_\mu)_\A,
\label{eqjap35}
\eea
and the superspace constraints following from \bref{U1A}
\bea
F_{\A\B}&=&(\bD_\A \7\bA_\B+\bD_\B \7\bA_\A)-2{{i}}(C\G^\mu)_{\A\B}A_\mu=0
\label{Fmnab}\eea
have been used in  
\bref{Theom}.
The sector of our model covered by $ (x^\mu,p_\mu,\theta^\alpha, \z_\A,$ $
f_{\mu\nu},\tilde\lambda_\alpha, \rD)$ describes therefore
a massless superparticle minimally coupled
to the super U(1) gauge field.
Identifying the interaction term $\CL^I=\CA$ in \bref{Lag0} with the EM 
one-form superpotential,
the two-superform field strength $\CF=d\CA$ is, after
using the MC equations \bref{MCBS},
\bea
\CF&=&d\CA=
\frac12f_{\mu\nu}\,L^\mu\,L^\nu+{{i}}\blam_\A\,(\G_{\mu} L)^{\A}\,L^\mu\,
+\cdots
\label{eqjap36}
\eea
where the $\cdots$ terms are linear in the one forms $L_B, L_\Sig^\A,
L_Z^{\mu\nu}$ which vanish on shell.
 The  field strength components are the ones given in \bref{eqjap35}-
\bref{Fmnab}.

Our model describes the
coupling to a particular choice of $U(1)$ gauge superfield strength
$W_\alpha(x,\theta)$ in \bref{eqjap1.4},
which satisfies the standard superspace constraints for
 the SUSY gauge theories
  \cite{Sohnius:1985qm},
\bea
&F_{\A\B}= 0,\; \quad   \bF_{\mu\A}=\bW_\B(\G_\mu{)^\B}_{\A},\;&
\cr
&
\bD_\A\bW_\B
=  -\frac{i}2(C\G^{\mu\nu})_{\A\B} \bF_{\mu\nu},\quad
\pa_\mu \bW_\B(\G^\mu{)^\B}_{\A}=0.
\label{solsuper}\eea
It is known (see e.g. \cite{Rocek:1989sh}) 
that the coupling of the N=1 superparticle to the gauge superfield strength
$\bW_\alpha(x,\theta)$ satisfying the constraints \bref{solsuper}
leads to a $\kappa$-invariant interaction.
Actually our system is not only invariant under the  global \SM\, 
symmetries but also invariant under $\tau$ reparametrization and the
$\kappa$ symmetries.

%%%%%%%%%%%%%%%%%%%%%%%%%%%%%%%%%%%%%%%%%%

\section*{Conclusions}

In this paper we found supersymmetric extensions of the Maxwell algebra
and proposed a  $\kappa$ invariant superparticle model \bref{Lag0}  with the
\SM\, symmetries. It couples minimally to a constant $U(1)$ gauge
superfield strength satisfying the superspace constraints
(see \bref{solsuper}).
It gives a new geometric framework for a superspace filled with a
uniform SUSY gauge field by generalizing the known non-supersymmetric one
with Maxwell symmetries. Because supersymmetries have critical importance
in current fundamental interaction theories (e.g. string/M-theory), 
we hope such a generalization will be useful in this context, in particular
in the interpretation of fermionic backgrounds. 

The \SM \,algebra is realized if we regard the variables
$ (f_{\mu\nu}, 
\tilde\lambda_\alpha, D$) as dynamical ones. 
In the Hamiltonian formulation of our model \bref{Lag0} they become the generators 
$(Z_{\mu\nu},\Sigma_\A,B)$ of the \SM\, symmetries.
Note that by taking a fixed solution for $ (f_{\mu\nu} , \tilde\lambda_\alpha,
D$) the {\SM}  symmetry is spontaneously broken to smaller ones
similarly as in the bosonic case \cite{Bacry:1970ye}.
The evolution of the coordinates 
 $(\phi^{\mu\nu}, \phi^\alpha,$ $\phiD)$ are described by Eq.\bref{phieom}
with their solutions 
determined by the trajectories in the ``physical'' subspace ($x_\mu,
\theta_\alpha , f_{\mu\nu} , \tilde\lambda_\alpha , D$).
It will be interesting to find some physical interpretation for the new coordinates
$(\phi^{\mu\nu}, \phi^\alpha, \phiD)$ and their dynamical roles.
For the bosonic Maxwell case it has been suggested \cite{Bonanos:2008ez}  that
$\phi^{\mu\nu}$ describes the magnetic moment
of a distribution of charged particles with center-of-mass position $x^\mu$.

The  \SM \,algebra $\CG$ introduced in this paper is a minimal
superextension of the Maxwell algebra. It can be considered 
as an enlargement of the Green algebra \cite{Green:1989nn}  by adding the tensorial central charges $Z_{\mu\nu}$.
In the Green algebra the 
spinorial generators $\Sig_\alpha$ are central (compare with \bref{15}).
We have considered also its central extension ${\widetilde\CG}$
and the enlargement $\CG_5$ by means of the chiral generator
$B_5$.
The {\SM}\, algebra $\CG$ can be embedded into larger superalgebras, 
in particular in the known Bergshoeff-Sezgin (BS) p-brane
algebra \cite{Bergshoeff:1995hm}.
 Thus one can introduce a corresponding BS-invariant superparticle model 
 with the interaction Lagrangian generalizing
\bref{Lag021} and gauge superpotentials ${A_\mu^{BS}}$, ${A_\alpha^{BS}}$ depending in a unique way on the BS supergroup coordinates. 
Using the coset with Lorentz stability group we find that the corresponding 
superfield strength  $F^{BS}$'s do not satisfy the superspace constraints 
\bref{solsuper},  i.e. the BS superparticle dynamics is not 
$\kappa$ symmetric. 
The origin of the non-invariance is the appearance of $Z_{\mu\nu}$ in the
$\{Q,Q\}$ anti-commutator 
resulting in $F_{\A\B}\neq0$ which violates the SUSY constraint
\bref{solsuper} (cf.\bref{eqjap36}).
 We note also  that Soroka and Soroka proposed in
 \cite{Soroka:2004fj}\cite{Soroka:2006aj} a nonstandard supersymmetrization of Maxwell algebra, without the translation generators in the basic anticommutator $\{Q,Q \}$; 
moreover in \cite{Soroka:2006aj} there is presented 
some superextension  of  k-deformed Maxwell algebra ($k>0$ of \cite{Gomis:2009vm}).

Our geometric scheme introduces additional degrees of freedom, describing
uniform gauge field strengths  in space and superspace leading to 
uniform constant energy density.
These global degrees of freedom are  dynamical, i.e.,  our model
provides a  framework in which the cosmological constant could be considered 
as a dynamical quantity. 
Recently many papers propose new types of dynamics 
to explain the dark energy phenomenon (see e.g.\cite{Copeland:2006wr}) 
as well as the dynamical  role  of the
cosmological constant  (see e.g.\cite{Caldwell:2005tm}\cite{Mukohyama:2003nw}).
Because at present these issues are of fundamental importance,
the developments in this paper should find some important applications.
\vs

 {\bf Acknowledgments}

We acknowledge discussions with Jorge Alfaro. We also acknowledge 
financial support from projects FPA2007-66665-C02-01, 2009SGR502, 
Polish Ministry of Science and High Education grant NN202 318534 and 
Consolider CPANCSD2007-00042.

\end{document}